# MIMETIZATION OF THE ELASTIC PROPERTIES OF CANCELLOUS BONE VIA A PARAMETERIZED CELLULAR MATERIAL

Lucas Colabella[a], Adrián P. Cisilino[a], Guillaume Häiat[b] and Piotr Kowalczyk[c]

[a] INTEMA- Faculty of Engineering, National University of Mar del Plata – CONICET
Av. Juan B. Justo 4302, Mar del Plata B7608FDQ, Argentina

[b] CNRS – Laboratoire Modélisation et Simulation Multiéchelle, UMRS CNRS 8208, 61 avenue du gal de Gaulle, 94010 Creteil, France

[c] Institute of Fundamental Technological Research, Polish Academy of Sciences, Pawinskiego 5B, 02-106 Warsaw, Poland

**ABSTRACT**

Bone tissue mechanical properties and trabecular microarchitecture are the main factors that determine the biomechanical properties of cancellous bone. Artificial cancellous microstructures, typically described by a reduced number of geometrical parameters, can be designed to obtain a mechanical behavior mimicking that of natural bone. In this work, we assess the ability of the parameterized microstructure introduced by Kowalczyk (2006) to mimic the elastic response of cancellous bone. Artificial microstructures are compared with actual bone samples in terms of elasticity matrices and their symmetry classes. The capability of the parameterized microstructure to combine the dominant isotropic, hexagonal, tetragonal and orthorhombic symmetry classes in the proportions present in the cancellous bone is shown. Based on this finding, two optimization approaches are devised to find the geometrical parameters of the artificial microstructure that better mimics the elastic response of a target natural bone specimen: a Sequential Quadratic Programming algorithm that minimizes the norm of the difference between the elasticity matrices, and a Pattern Search algorithm that minimizes the difference between the symmetry class decompositions. The pattern search approach is found to produce the best results. The performance of the method is demonstrated via analyses for 146 bone samples.

## 1 INTRODUCTION

Bones are hierarchical bio-composite materials with a complex multiscale structural geometry (Carretta et al. 2013). Bone tissue is arranged either in a compact pattern (cortical bone) or a spongy pattern (cancellous bone). Cancellous bone can be found in vertebral bodies and at the epiphyses of long bones. In the vertebral body, it is the main load bearing structure, where as in the appendicular skeleton, it transfers mechanical loads from the articular surface to cortical bone. Furthermore, trabecular bone quality is an important determinant of the overall bone strength and affects fracture risk. To better understand the mechanics of cancellous bone is of interest for the diagnosis of bone diseases (osteoporosis), the evaluation of the risk of fracture, and for the design of artificial bone (Cowin 2001).

Cancellous bone can be assimilated to a composite material with hierarchical structure. In a bottom-up description, the structure starts in the nanoscale (mineralized collagen fibril) and moves up to the sub-microscale (single lamella), the microscale (single trabecula), and mesoscale (trabecular bone) levels. Trabeculae are organized into a three-dimensional lattice oriented mainly along the lines of stress, which forms a stiff and ductile structure that provides the framework for the soft bone marrow filling the intertrabecular spaces. Trabeculae consist of a nanometric extracellular matrix that incorporates hydroxyapatite, the bone mineral that provides bones rigidity (Sansalone et al. 2010, 2012), and collagen, an elastic protein which improves fracture resistance (Keaveny et al. 2001).

Bone tissue mechanical properties and trabecular architecture are the main factors determining the mechanical properties of cancellous bone, which show a high dependency on species, anatomic site, age and size of the sample (Fritsch and Hellmich 2007; Parkinson and Fazzalari 2013). The small dimensions of the trabeculae (of the order from tens to a cent of microns) hinder their mechanical characterization at tissue-level. In recent years, nano-indentation has provided the means for the direct measurement of the elastic properties of trabecular bone tissue (a complete review of the available techniques, many of them indirect, can be found in the recent paper by Oftadeh et al. (2015)). By means of high resolution nano-indentation, Brennan et al. (2009) studied the tissue property variations within a single trabecula; they found that Young´s modulus and hardness increase towards the core of the trabeculae. Despite this findings, it is the common assumption that the mechanical inhomogeneity and anisotropy of bone tissue has a minor impact on the apparent properties of cancellous bone and, consequently, it can be approximated by an isotropic tissue modulus (Kabel et al. 1999a, b).

Different experiments have shown that linear elasticity can predict the behavior of cancellous bone (Keaveny et al. 1994). The trabecular architecture determines the elastic anisotropy of cancellous bone, which can be described by a fourth rank elastic tensor, $\mathbb{C}$, which linearly relates stress and strains as $\sigma = \mathbb{C} \cdot \varepsilon$. The elastic tensor is determined in its most general form by 81 components. Cancellous bone is generally assumed to behave as an orthotropic structure in the mesoscale, with three planes of symmetry, what requires of only nine independent components to fully describe the elastic behavior of the structure (Yang et al. 1998).

Relationships between elastic properties and structural parameters of cancellous bone have been proposed in the literature. Apparent density, $\rho_{app}$, which is the product of bone volume fraction, $BV/TV$, and of bone tissue density, $\rho$, is the primary component affecting the mechanical properties of cancellous bone (Oftadeh et al. 2015). Several studies have proposed linear and exponential forms that express the Young's modulus as a function of $\rho_{app}$, showing a high correlation with the experiments. However, these correlations exhibit limitations: the error in the elastic moduli can be up to 53% at certain volume fractions (van Rietbergen et al. 1999) and they cannot individually predict the elastic properties of trabecular bone in different anatomic sites and species (Helgason et al. 2008). Similarly, cancellous bone elastic properties can be correlated to the fabric tensor, which is a descriptor of the anisotropy of trabecular bone (Gross et al. 2013; Maquer et al. 2015). Another approach is to use large-scale finite element (FE) analysis of microstructural models built from micro-computed tomographic (CT) scans of real bone specimens. Finite element analyses can solve some of the drawbacks of the experimental techniques, since FE models can be subjected without restrictions to the load conditions needed to evaluate the anisotropic behavior of the microstructure. Moreover, the combination of finite element analysis (FEA) and modern in vivo high-resolution peripheral quantitative CT scanners, allow for the analysis of bone in-vivo in the peripheral skeleton (van Rietbergen and Ito 2015). FEA has been applied to large sets of data to find the orthotropic components of cancellous bone (Kabel et al. 1999a, 1999b), which show that there are strong correlations between $BV/TV$ and elastic and shear moduli, whereas this correlation is weak for the Poisson's ratio.

Another approach is to use parametric models of trabecular bone, which consist in artificial microstructures formed by plates and rods. Artificial microstructures may be criticized for being somewhat unrealistic. However, their main advantage is that the mesoscopic properties characterizing such microstructures can be expressed as explicit continuous functions of some well-defined geometrical parameters. Moreover, it has been found that models based on local morphometry, composed of individual rods and plates, help improving the understanding of local structural changes in the determination of bone stiffness (Stauber and Müller, 2006a, 2006b). Explicit relations between geometrical parameters and mesoscopic properties are crucial for modeling the microstructure evolution at the large scale - it allows to formulate the problem as merely the evolution of a set of scalar variables, which is much more efficient in terms of computational cost than the analysis of the geometric evolution of certain components of micro-CT-based actual bone microstructures. Examples in this sense are the artificial trabeculae developed by Kowalczyk (2006) and Dagan et al. (2004), which have been successfully employed in the modeling of long-term changes in morphological and mechanical properties of trabecular bone in the proximal femur, see Be'ery-Lipperman and Gefen (2005) and (Kowalczyk 2010), respectively. Artificial trabecular microstructures have been also used for the developing of continuum models to describe the trabecular-bone stress-strain response (Goda et al. 2014; Goda and Ganghoffer 2015a) and multiaxial yield and failure criteria (Goda and Ganghoffer 2015b) by means of homogenization analyses. In turn, homogenized properties of parametric models of trabecular bone can be integrated into multiscale optimization methods for the design of bone substitutes and natural micro-scaffolds (Hollister 2005) for tissue engineering. Such hierarchical structures can be fabricated by means of emerging 3D printing technologies (Bose et al. 2013; Wang et al. 2016).

In this work the parameterized cancellous microstructures introduced by Kowalczyk (2006) are analyzed in terms of their ability to mimic the elastic response of natural cancellous bone. Artificial microstructures are compared with actual bone samples in terms of their symmetry classes and their elasticity matrices represented in terms of the geometry parameters. Two optimization approaches are proposed in this paper to design the parameterized microstructure that better mimics the elastic response of a target natural bone specimen: a Sequential Quadratic Programming algorithm that minimizes the norm of the difference between the elasticity matrices, and a Pattern Search algorithm that minimizes the difference between the symmetry class decompositions. Both approaches use the geometry parameters as design variables.

## 2 ELASTIC PROPERTIES OF CANCELLOUS BONE

### 2.1 Experimental data

Two sources of experimental data are used: the database for human cancellous-bone specimens by Kabel et al. (1999a, 1999b), and five bovine femoral bone specimens that are examined as part of this work.

The database by Kabel et al. (1999a, 1999b) provides the entire set of anisotropic elastic constants of 141 human cancellous-bone specimens of vertebral body, calcaneus, proximal tibia and distal femur. Specimen bone volume-to-total volume ratios cover the range $5\% \leq BV/TV \leq 35\%$. The elastic constants are the results of finite element (FE) homogenization analyses performed on computer reconstructions of the specimen microarchitectures. Linear elastic and isotropic material properties were specified for the bone tissue, with a Young's modulus of $E = 1\ GPa$ and Poisson's ratio $\nu = 0.3$, so the homogenized results can be scaled for any value of the tissue modulus. The specimen imaging and homogenization procedures are fully described in Kabel et al. (1999b).

The five bovine samples were obtained from femoral bones. They were X-ray scanned using micro computed tomography (μCT) with a resolution of 17.7 $\mu m$. CT images were processed with BoneJ (Doube et al. 2010) to obtain geometrical data over the Regions of Interest (ROI). The results for bone volume fractions, trabecular thicknesses and trabecular spacings are reported in Table 1. The elastic modulus of the bone tissue was measured via microindentation tests using the Oliver and Pharr (1992) method. Specimens were microindented in a TI 900 Triboindenter (Hysitron, MN, USA) using a Vickers diamond indenter. The maximum indentation load was set to 1500 $mN$, which was held constant for 45 $s$ in order to minimize creep effects. The loading and unloading rates were set to 200 $mN/s$ and 100 $mN/s$ respectively. Eight indentations were performed on each sample. The resulting Young's modulus was $E_b = 7.93 \pm 0.86\ GPa$. A Poisson's ratio $\nu = 0.3$ was assumed. The elastic homogenization analyses were performed using the Fast Fourier Transform method with models built after the μCT geometry data. The procedure is described in Colabella et al. (2017). The resulting elasticity matrices $C_b$ for the five specimens are reported in Appendix 1.

| Specimen | ROI dimensions [$mm$] | Pixel size [$\mu m$] | BV/TV [%] | Trabecular thickness, $t$ [$\mu m$] | Trabecular spacing, $s$ [$\mu m$] | Normalized trabecular thickness, $t/(t+s)$ |
|---|---|---|---|---|---|---|
| 1 | 8.7x9.6x9.5 | 17.7 | 25 | 157 | 491 | 0.24 |
| 2 | 9.9x10.3x9.8 | 17.7 | 38 | 241 | 471 | 0.34 |
| 3 | 9.8x10.1x7.7 | 17.7 | 30 | 180 | 621 | 0.22 |
| 4 | 9.6x9.2x8.9 | 17.7 | 20 | 180 | 757 | 0.19 |
| 5 | 9.7x9.5x7.6 | 17.7 | 21 | 145 | 573 | 0.20 |

*Table 1: Geometrical data of the bovine samples*

## 2.2 Elastic symmetry analyses

We propose here to study the elastic symmetries of the samples by decomposing their elasticity tensors into sums of orthogonal tensors belonging to the different symmetry classes. We use for this purpose the method by Browaeys and Chevrot (2004). This method relies on the following vectorial description of the elasticity tensor,

$$\boldsymbol{X} = (C_{11}, C_{22}, C_{33}, \sqrt{2}C_{23}, \sqrt{2}C_{13}, \sqrt{2}C_{12}, 2C_{44}, 2C_{55}, 2C_{66}, 2C_{14}, 2C_{25}, 2C_{36}, \\ 2C_{34}, 2C_{15}, 2C_{26}, 2C_{24}, 2C_{35}, 2C_{16}, 2\sqrt{2}C_{56}, 2\sqrt{2}C_{46}, 2\sqrt{2}C_{45}),$$

(1)

where $C_{ij}$ are the components of the elastic tensor $\mathbb{C}$ in the Voigt notation. The normalization factors in the above expression are included so that the Euclidean norm of an arbitrary elastic tensor $\mathbb{C}$ and its associated elastic vector $\boldsymbol{X}$ are identical.

The vector description of the elastic tensor possesses the property that any symmetry class constitutes a subspace of a class of lower symmetry and an orthogonal projection on this subspace removes the lower symmetry part, see Table 2. Thus, when expressed in the so-called symmetry Cartesian coordinate system (see Cowin and Mehrabadi, 1987), $\boldsymbol{X}$ can be decomposed by a cascade of projections into a sum of vectors belonging to the symmetry classes triclinic, monoclinic, orthorhombic, tetragonal, hexagonal and isotropic:

$$\boldsymbol{X} = \boldsymbol{X}_{tri} + \boldsymbol{X}_{mon} + \boldsymbol{X}_{ort} + \boldsymbol{X}_{tet} + \boldsymbol{X}_{hex} + \boldsymbol{X}_{iso}, \qquad (2)$$

The different elastic symmetry parts can be presented as fractions of the Euclidian norm of the elasticity vector $\|\boldsymbol{X}\|$ as follows:

$$\begin{aligned}
c_{iso} &= 1 - \frac{\|\boldsymbol{X}_{tri} + \boldsymbol{X}_{mon} + \boldsymbol{X}_{ort} + \boldsymbol{X}_{tet} + \boldsymbol{X}_{hex}\|}{\|\boldsymbol{X}\|} \\
c_{hex} &= 1 - c_{iso} - \frac{\|\boldsymbol{X}_{tri} + \boldsymbol{X}_{mon} + \boldsymbol{X}_{ort} + \boldsymbol{X}_{tet}\|}{\|\boldsymbol{X}\|} \\
c_{tet} &= 1 - c_{hex} - c_{iso} - \frac{\|\boldsymbol{X}_{tri} + \boldsymbol{X}_{mon} + \boldsymbol{X}_{ort}\|}{\|\boldsymbol{X}\|} \\
c_{ort} &= 1 - c_{tet} - c_{hex} - c_{iso} - \frac{\|\boldsymbol{X}_{tri} + \boldsymbol{X}_{mon}\|}{\|\boldsymbol{X}\|} \\
c_{mon} &= 1 - c_{ort} - c_{tet} - c_{hex} - c_{iso} - \frac{\|\boldsymbol{X}_{tri}\|}{\|\boldsymbol{X}\|} \\
c_{tri} &= 1 - c_{mon} - c_{ort} - c_{tet} - c_{hex} - c_{iso}
\end{aligned} \qquad (3)$$

so that

$$c_{iso} + c_{hex} + c_{tet} + c_{ort} + c_{mon} + c_{tri} = 1. \qquad (4)$$

Computations for the determination of the symmetry Cartesian coordinate system, the transformations into vector forms, the symmetry decompositions and normalizations were performed using the Matlab Seismic Anisotropy Toolkit (MSAT) by Walker and Wookey (2012).

| Symmetry class | Planes of symmetry | Dimension |
|---|---|---|
| Triclinic | 0 | 21 |
| Monoclinic | 1 | 13 |
| Orthorhombic | 3 | 9 |
| Tetragonal | 5 | 6 |
| Hexagonal | 7 | 5 |
| Isotropic | ∞ | 2 |

*Table 2: Number of planes of symmetry and dimension of the subspaces. Note that the dimension number corresponds to the number of distinct components of the elastic tensor.*

Figure 1 presents the results for the symmetry class decompositions of the 141 human bone samples in Kabel et al. (1999a, 1999b). The cumulative decompositions in (4) are presented as functions of the sample $BV/TV$. Table 3 summarizes the extreme values for the symmetry classes. It is observed from Figure 1 that the isotropic class accounts for the most significant fraction of the elasticity matrices over the complete $BV/TV$ range. Although its wide dispersion, the mean value of the isotropic fraction increases linearly with $BV/TV$, from $\overline{c_{iso}} = 0.49$ for $BV/TV = 5\%$ to $\overline{c_{iso}} = 0.69$ for $BV/TV = 35\%$. Its standard deviation is $SD_{iso} = 0.12$. The second relevant fraction is for the hexagonal class. Conversely to the isotropic class, the mean value of the hexagonal class decreases linearly with $BV/TV$. The isotropic and hexagonal classes behave such that they add a constant, $\overline{c_{iso}} + \overline{c_{hex}} = \cong 0.83$, with a standard deviation $SD_{iso+hex} = 0.07$. The tetragonal class fraction is marginal; around 1% for nearly 98% of the samples. The orthorhombic class presents a wide dispersion, but its mean value is nearly constant $\overline{c_{ort}} \cong 0.10$. The orthotropic symmetry,

$$c_{ortho} = c_{ort} + c_{tet} + c_{hex} + c_{iso}, \qquad (5)$$

presents a constant average value $\overline{c_{ortho}} = 0.93$ with $SD_{ortho} = 0.04$. This last result is consistent with the observation by Yang et al. (1998), who found that $\mathbb{C}$ matrices present orthotropic symmetry with a 95% confidence level.

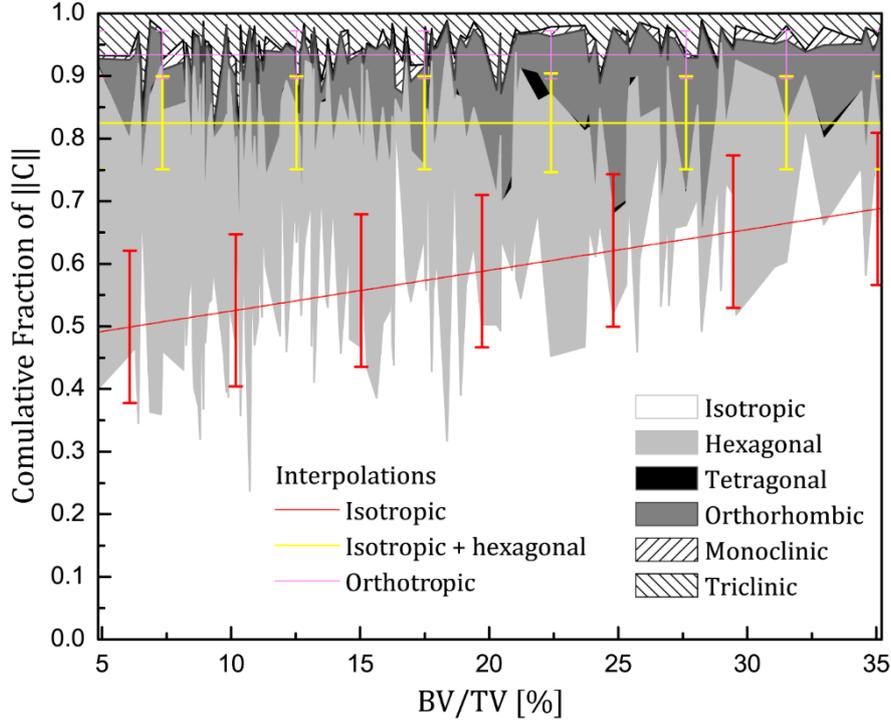

*Figure 1: Symmetry class decomposition of the elasticity matrices of the 141 human-bone specimens reported by Kabel et al. (1999a, 1999b). Error bars indicate the standard deviations from the interpolated mean values.*

| Symmetry class | Human Samples Kabel et al (1999) | | Bovine Samples (this work) | | Parameterized Kowalczyk (2006) | |
|---|---|---|---|---|---|---|
| | Min | Max | Min | Max | Min | Max |
| $c_{iso}$ | 0.24 | 0.84 | 0.50 | 0.79 | 0.36 | 1.00 |
| $c_{hex}$ | 0.02 | 0.65 | 0.04 | 0.20 | 0.00 | 0.49 |
| $c_{iso} + c_{hex}$ | 0.56 | 0.95 | 0.66 | 0.89 | 0.46 | 1.00 |
| $c_{tet}$ | 0.00 | 0.05 | 0.00 | 0.01 | 0.00 | 0.15 |
| $c_{ort}$ | 0.00 | 0.33 | 0.02 | 0.13 | 0.00 | 0.50 |
| $c_{ortho}$ | 0.81 | 0.99 | 0.79 | 0.91 | 1.00 | 1.00 |
| $c_{mon}$ | 0.00 | 0.08 | 0.00 | 0.04 | - | - |
| $c_{tri}$ | 0.01 | 0.19 | 0.08 | 0.19 | - | - |

*Table 3: Extreme values of the symmetry classes of the natural and parameterized trabecular microstructures.*

The results for the symmetry class decompositions of bovine samples are in Figure 2. Extreme values for the symmetry classes are in Table 3. It is observed that although extreme values of the bovine symmetry classes are within those of the human data ranges, the sum of the isotropic and hexagonal classes is $0.66 \leq c_{hex} + c_{iso} \leq 0.89$, which has the lower limit outside the standard deviation of the human data. A similar behavior is observed for the orthotropic symmetry of the bovine samples, $0.79 \leq c_{ortho} \leq 0.91$.

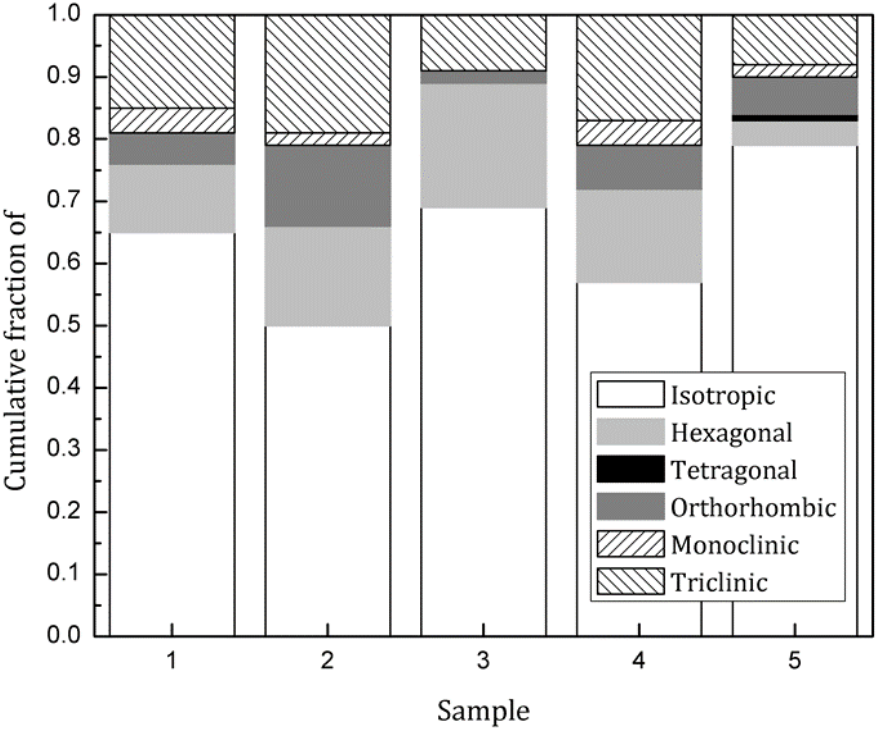

Figure 2: Symmetry class decomposition of the bovine samples.

# 3 DEVELOPMENT OF A MIMETIC CANCELLOUS BONE MICROSTRUCTURE

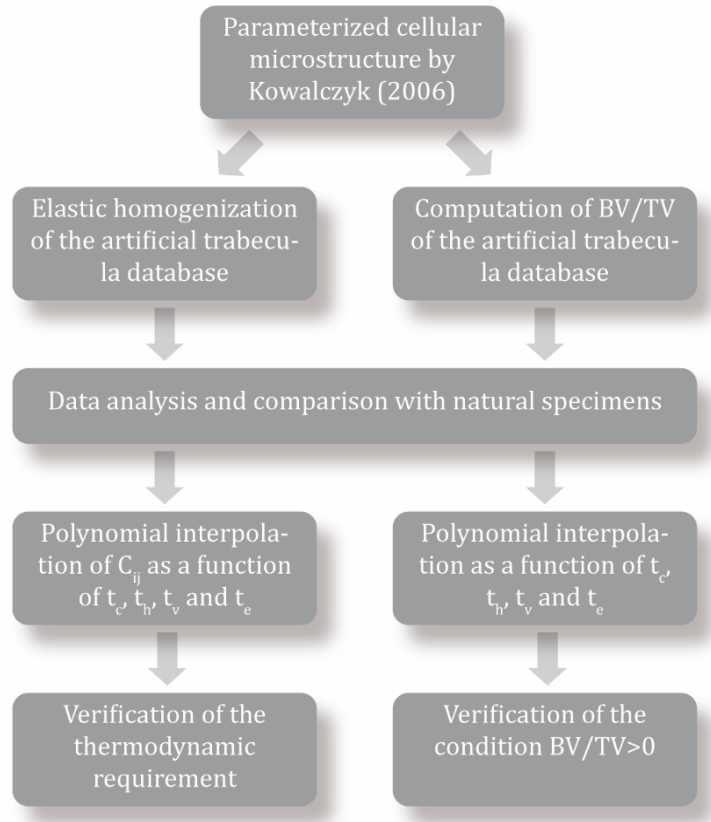

*Figure 3: Workflow of the development of a mimetic cancellous bone microstructure.*

The workflow in Figure 3 depicts the procedure for the development of the mimetic cancellous bone microstructure. The first step is the computation of the elasticity and BV/TV data of the parameterized cellular microarchitecture introduced by Kowalczyk (2006), on which the development is based. The capabilities of the parameterized microstructure to mimic natural bone is assessed through the comparison and correlation of the elastic and BV/TV data with that of the natural specimens. Next, the elastic and BV/TV data are interpolated as functions of the geometrical parameters. These interpolations will play a key role in the implementation of the optimization algorithms in Section 4. Finally, the polynomial interpolations are checked for consistency. The details of this procedure are given next.

## 3.1 The parameterized cellular microstructure

The parameterized cellular microarchitecture introduced by Kowalczyk (2006) is shown in Figure 4. It consists in a repeatable cell that is inscribed into a space-filling dodecahedron, so it can be arranged in rows and layers to completely fill the 3-D space. The geometry of the cell is described by Bezier curves and corresponding surface patches. Surface transitions between neighboring cells are smooth. Shaded areas denote trabecular surface while the hatched areas are the cross-sections at which the cell is "stuck" to identical neighboring cells.

The repeatable geometry is described in terms of four geometrical parameters: $t_c$, $t_h$ and $t_v$, which define proportions between trabecular plate widths and thicknesses to produce transversely isotropic microstructures in the $x_1 - x_2$ plane; and $t_e$, which scales it in the $x_1$ direction to produce fully orthotropic microstructures. Parameters $t_c$, $t_h$ and $t_v$ are non-dimensional as they are understood as fractions of the corresponding cell dimensions and may take values between 0 and 1. In order to produce feasible geometries, they must comply with the restrictions

$$t_h \geq t_c, \qquad t_v \geq t_c. \tag{6}$$

Parameter values can be set to produce microstructures with solid volume fractions in the range $0 < BV/TV < 100\%$.

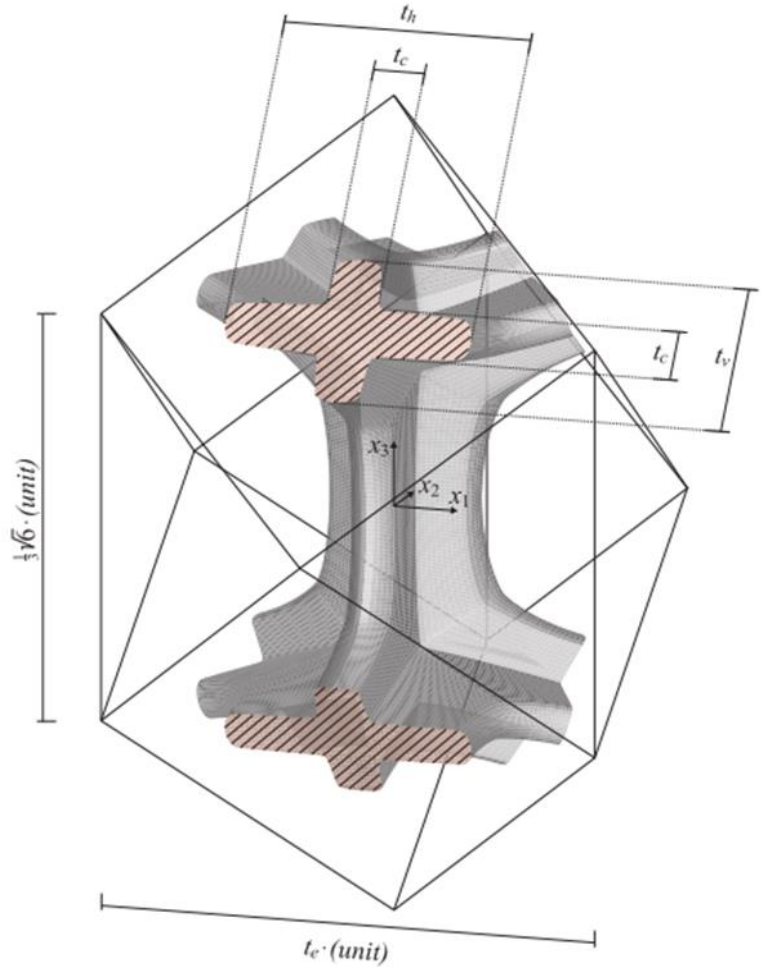

*Figure 4: Geometry of a repeatable cell.*

### 3.2 Homogenization analysis and comparison with natural specimens

Kowalczyk (2006) used FE homogenization to compute the elasticity matrices for a broad set of microstructures given by $(t_c, t_h, t_v, t_e)$ quadruplets on the domains $t_c \in (0,1)$, $t_h \in [t_c, 1)$ and $t_v \in [t_c, 1)$ in increments of 0.05, $t_e \in [0.6, 1.4]$ in increments of 0.2, and elastic properties $E = 1$ GPa and $\nu = 0.3$. He studied in detail the dependency of the resultant effective elastic constants with $BV/TV$, and tabulated functions for the elastic constants in terms of morphometric parameters: mean intercept length (MIL), volume orientation (VO) and star length distribution (SLD). From the comparison of the effective elastic constants to those of the natural samples by Kabel et al. (1999a, 1999b) he showed that individual ranges of $E_{ij}$, $G_{ij}$ and $\nu_{ij}$ of the parameterized microstructures are always wider than those of the natural specimens for every $BV/TV$ value.

In what follows, the above analysis is extended to assess the capability of the parameterized microstructure to mimic the natural elastic symmetries. To this end, the database of elastic constants for the parameterized microstructure was refined, such that the four geometric parameters were varied in increments of 0.05. The FE homogenization procedure of Kowalczyk (2006) was used for the computation of the effective elastic constants. The repeatable cells that result for each combination of the geometric parameters were discretized with 8-node linear brick elements and appropriate boundary conditions that ensure fitting of all deformed neighboring cells to each other were specified. Six load cases were considered for each cell: pure stretching in three orthogonal directions $x_1, x_2, x_3$ (see Figure 4) and pure shear in three orthogonal planes (normal to the three directions), which were specified in terms of the displacement fields. Reaction forces were measured for each case and used to compute the elements of the elasticity matrix. Thus, the construction of the database consisted of 41,990 homogenization analysis that involved the solution of approximately 250,000 finite-element models altogether.

Stiffness matrices of the parameterized microstructures are decomposed into their symmetry classes using the same procedure introduced earlier for the natural specimens. The extreme values attained by the symmetry classes are reported in Table 3. It is found that the parameterized microstructure covers the complete extents of the tetrahedral and orthorhombic classes of the natural specimens. On the other hand, it fails to cover the lowermost values of the isotropic class, $0.24 \leq c_{iso} < 0.36$, and the uppermost values of the hexagonal class, $0.49 > c_{hex} \geq 0.65$ for the human specimens. However, it is worth noting that only a few samples lie within the excluded ranges: 5 samples have $c_{iso} < 0.36$ and 2 samples have $c_{hex} > 0.49$, i.e., less than 5% of the 141 samples in the database. Symmetry classes of the bovine samples lie always within the extents of the parameterized microstructure.

The capability of the parameterized microstructure to mimic the elastic behavior is further assessed in terms of the $BV/TV$. Figure 5 shows the symmetry classes of the parameterized microstructures for the range of $BV/TV$ of the natural samples; these are 17,522 data points (in order to keep the figure clear not all data points were plot). Results are presented for $c_{iso}$, $c_{iso} + c_{hex}$ and $c_{ortho}$ in subfigures ($a$), ($b$) and ($c$), respectively. The corresponding data for the natural samples are also shown in the figures: gray areas indicate the standard deviation of the human samples (see Figure 1) while the square marks are the values of the bovine samples (see Figure 1). Figure 5($a$) shows that with the only exception of the lowermost values, i.e., for $5\% \lesssim BV/TV \lesssim 7\%$, the parameterized microstructure is able to mimic the isotropic class of the natural trabeculae. Regarding $c_{iso} + c_{hex}$, Figure 5($b$) shows that the parameterized microstructure completely encompasses the data of the natural samples (the gray area is hardly visible behind the symbols). Finally, the results for the orthotropic symmetry in Figure 5($c$) show that, consistently with its geometric definition, the parameterized microstructure shows $c_{ortho} = 1$ over the complete $BV/TV$ range, which results in a consistent overestimation of the orthotropic symmetry by the parameterized microstructure, the mean value and standard deviation of which is $6.6 \pm 3.8\%$ with respect to the human samples. Regarding the bovine samples, the overestimation ranges from 9% to 19%.

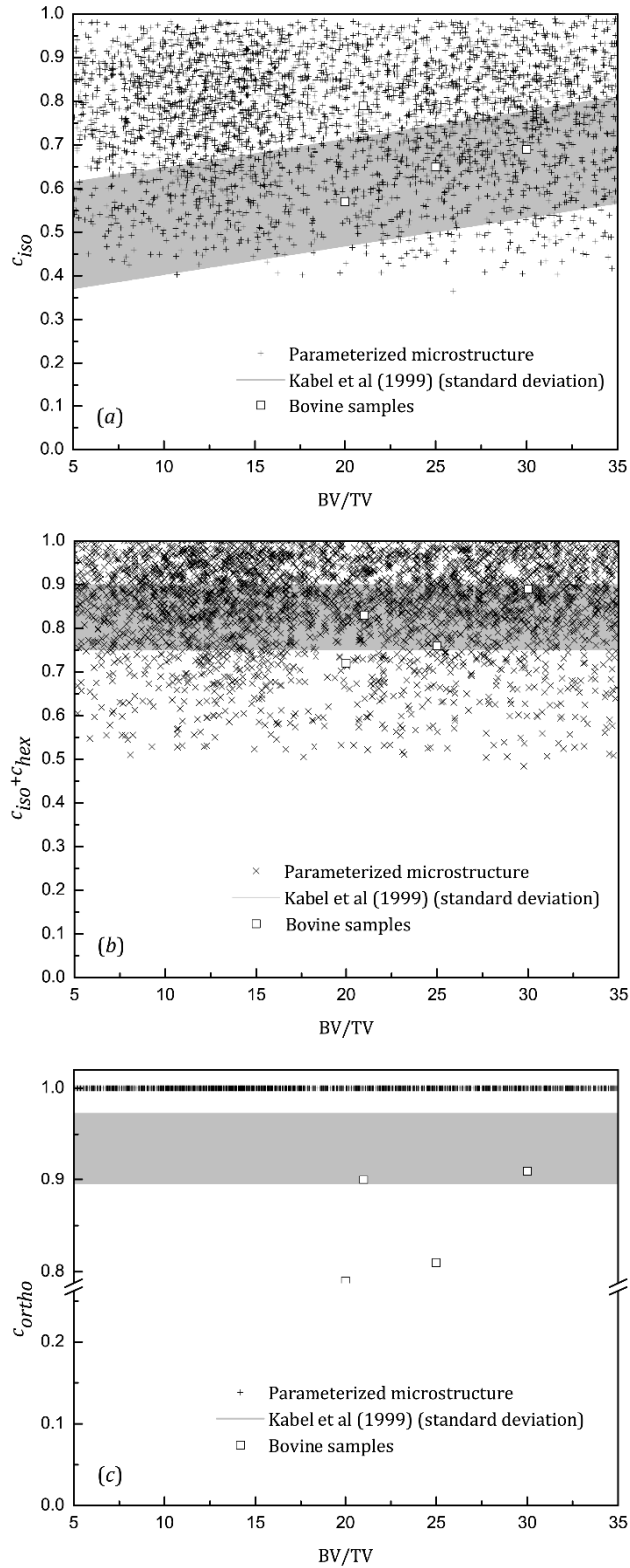

*Figure 5: Comparison of elastic symmetry classes of the natural and the parameterized trabecular microstructures: (a) isotropic class, (b) isotropic + hexagonal classes, (c) orthotropy.*

## 3.3 Polynomial interpolation

The discrete elastic-constant and symmetry-class data of the parameterized microstructures were examined to investigate their functionalities with the geometric parameters.

It was observed that coefficients of the stiffness tensor $\mathbb{C}$ behave as continuous and smooth functions of $t_c$, $t_h$, $t_v$ and $t_e$ over the complete range, and that in general, the $C_{ij}$ rise with the increment of the geometrical parameters. As examples, Figure 6 illustrates the behaviors of $C_{11}$ and $C_{12}$ as functions of $t_v$ and $t_c$ for $t_h = 0.6$ and $t_e = 1.2$.

On the other hand, symmetry classes showed to be, in some cases, discontinuous functions of the geometric parameters. Figure 7 depicts the changes of the symmetry classes associated to the variation of the elasticity coefficients given in Figure 6. It can be observed that, although $\mathbb{C}$ coefficients have a continuous variation, the hexagonal, tetragonal and orthorhombic symmetries present discontinuities.

Based on the above observations, the discrete elastic-constant data was used to interpolate an analytical expression for $\mathbb{C}(t_c, t_h, t_v, t_e)$. The nine non-zero coefficients, $C_{11}$, $C_{22}$, $C_{33}$, $C_{12}$, $C_{13}$, $C_{23}$, $C_{44}$, $C_{55}$ and $C_{66}$, were interpolated polynomially by means of least-square fitting. Polynomials of order 5 to 12 were used; the quality of the interpolations was assessed in terms of the coefficient of determination (R2), the root-mean-square error (RMSE) and the residual sum of squares (SSres). Table 4 reports the averaged for the interpolations of the 9 coefficients.

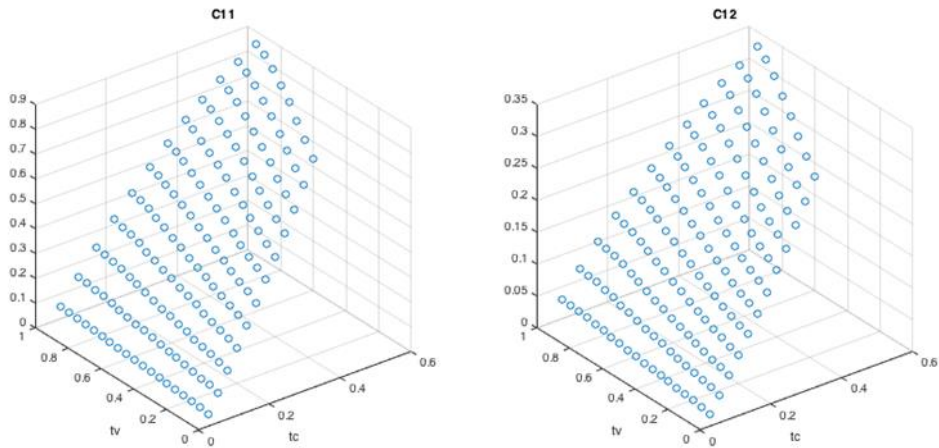

Figure 6: Parameterized-microstructure elasticity coefficients $C_{11}$ and $C_{12}$ as functions of $t_c$ and $t_v$ for $t_h = 0.6$ and $t_e = 1.2$.

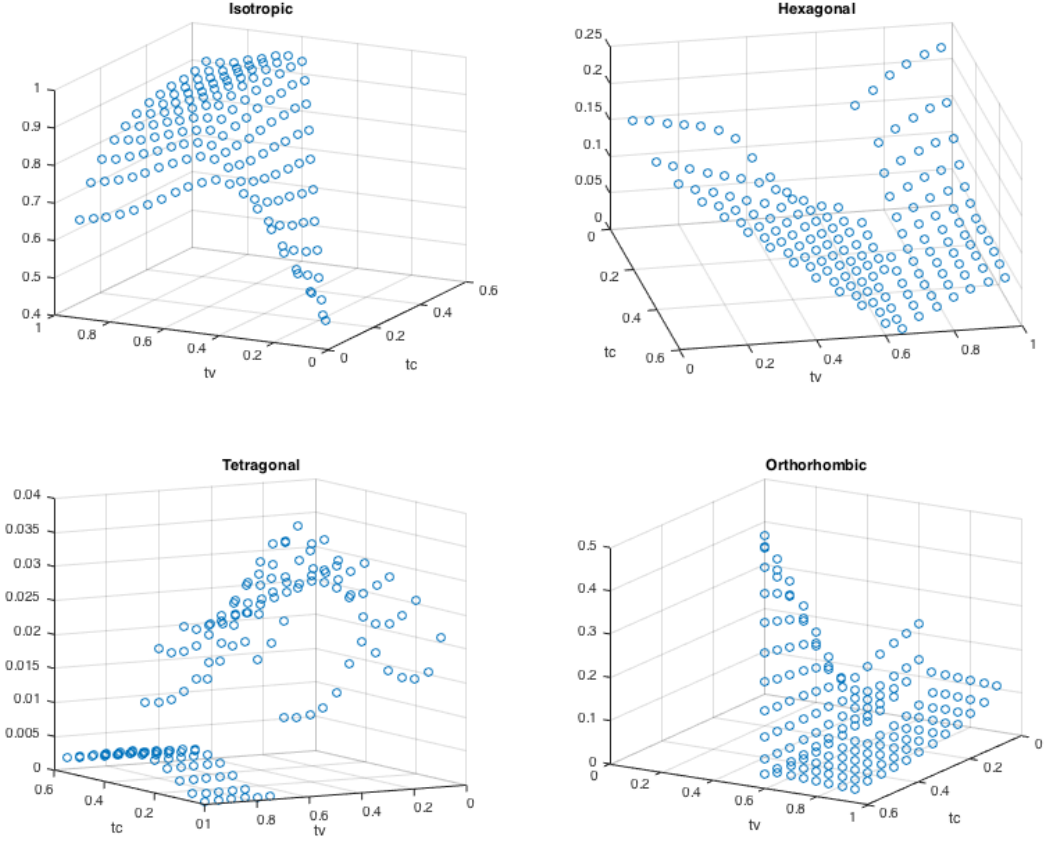

*Figure 7: Parameterized-microstructure symmetry classes as functions of $t_v$ and $t_c$ for $t_h = 0.6$ and $t_e = 1.2$.*

| Polynomial order | R2 | RMSE | SSres | Validity interval for $t_c$ |
|---|---|---|---|---|
| 5 | 0.9999848 | 0.000893 | 0.04095 | [0.16,0.95] |
| 6 | 0.9999965 | 0.000434 | 0.00979 | [0.15,0.95] |
| 7 | 0.9999986 | 0.000281 | 0.00410 | [0.09,0.95] |
| 8 | 0.9999991 | 0.000226 | 0.00265 | [0.11,0.95] |
| 9 | 0.9999993 | 0.000197 | 0.00204 | [0.10,0.95] |
| 10 | 0.9999994 | 0.000179 | 0.00167 | [0.06,0.95] |
| 11 | 0.9999995 | 0.000163 | 0.00139 | [0.09,0.95] |
| 12 | 0.9999996 | 0.000149 | 0.00115 | [0.06,0.95] |

*Table 4: Quality assessment of the polynomial interpolations of the elasticity coefficients in terms of the coefficient of determination ($R^2$), the root-mean-square error (RMSE) and the residual sum of squares (SSres).*

Since the $C_{ij}$ were interpolated separately, the thermodynamic requirement for $\mathbb{C}(t_c, t_h, t_v, t_e)$ was checked. The thermodynamic requirement (positive definiteness of strain energy) enforces the condition that the invariants of the elasticity matrix should be positive, or in other words, that both, $\mathbb{C}(t_c, t_h, t_v, t_e)$ and its inverse must be positive definite. These conditions were verified for the interpolated $\mathbb{C}(t_c, t_h, t_v, t_e)$ for quadruplets in the intervals $t_c \in [0.05, 0.95]$, $t_h \in [t_c, 0.95]$, $t_v \in [t_c, 0.95]$ and $t_e \in [0.6, 1.4]$ in increments of 0.01. The thermodynamic requirement was found valid on most of the interpolation range; validity intervals are reported in Table 4 in terms of $t_c$.

Based on the results in Table 4, the polynomial fittings of order 10 are selected. Polynomials of order 10 produce accurate and valid interpolations over a wide range of the parameter values. The validity interval $t_c \in [0.06, 0.95]$ allows for solid volume fractions $1\% \leq BV/TV \leq 99\%$.

Bone volume-to-total volume ratio data was also interpolated polynomially to obtain an analytical expression for $BV/TV\,(t_c, t_h, t_v, t_e)$. Polynomials of order 2 to 5 were explored; the results are summarized in Table 5. It can be observed that the quality of the interpolation consistently improves with the polynomial order. The reliability of the interpolations was verified by checking that they result in positive $BV/TV$ values. It was found that with the only exception of the interpolation of order 2, all the interpolations produce positive $BV/TV$ values for every combination of the geometrical parameters within their validity intervals. The polynomial fitting of order 5 is selected for the rest of the work.

| Polynomial order | $R^2$ | RMSE | SSres |
|---|---|---|---|
| 2 | 0.9987964 | 0.017454 | 12.79248 |
| 3 | 0.9999748 | 0.002527 | 0.26814 |
| 4 | 0.9999973 | 0.000821 | 0.02829 |
| 5 | 0.9999992 | 0.000442 | 0.00821 |

*Table 5: Quality assessment of polynomial interpolation of the BV/TV data.*

## 4 OPTIMIZATION

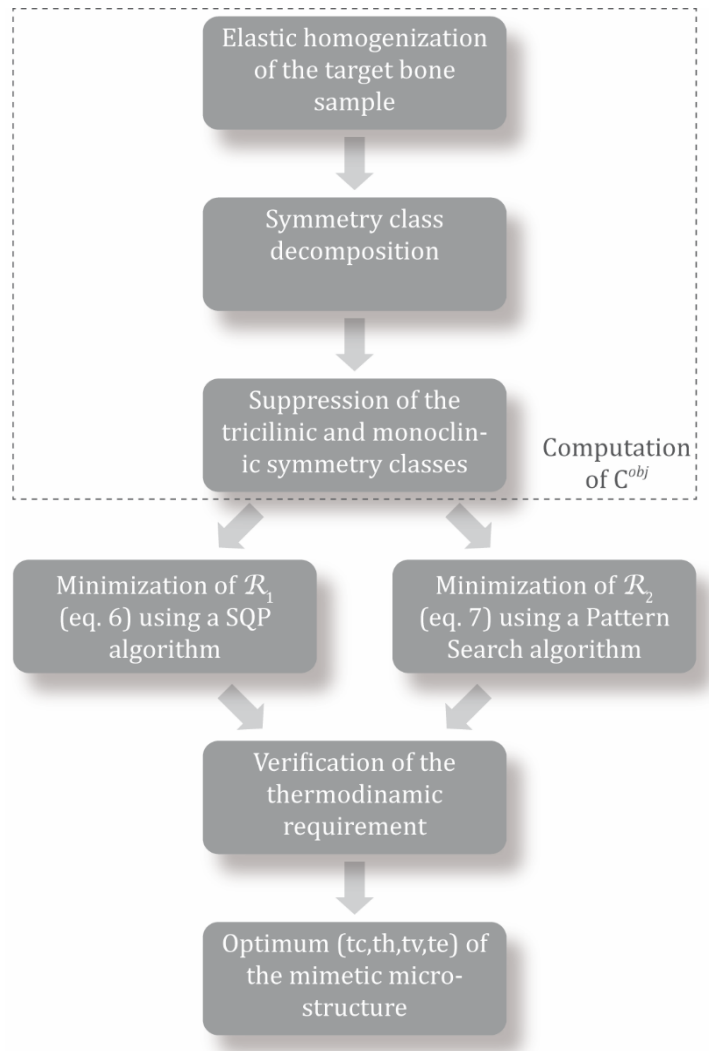

*Figure 8: Workflow of the optimization analysis.*

## 4.1 Problem statement

The optimization problem is to find the parameterized microstructure that better mimics the elastic response of a target natural bone specimen. The elastic equivalence among the microstructures is posed via two approaches: in terms of the elasticity matrix norm and in terms of its symmetry classes.

The workflow is illustrated in Figure 8. The analysis starts with the elastic homogenization of the target bone sample; the resultant elasticity matrix is decomposed into symmetry classes. The triclinic and monoclinic classes are suppressed to obtain the target elasticity matrix. Thus, the resultant $\mathbb{C}^{target}$ has orthotropic symmetry, and it is compatible with the artificial microstructure.

The problem posed in terms of the elasticity matrix norm consists in finding the quadruplet $t_c, t_h, t_v$ and $t_e$ that minimizes the norm of the difference between the $\mathbb{C}^{target}$ and $\mathbb{C}$:

$$\mathcal{R}_1 = \min \frac{\|\mathbb{C}^{target} - \mathbb{C}(t_c, t_h, t_v, t_e)\|}{\|\mathbb{C}^{target}\|}, \qquad (7)$$

The problem posed in terms of the symmetry classes consists in finding the quadruplet $t_c$, $t_h$, $t_v$ and $t_e$ that minimizes the overall difference between the symmetry class decompositions:

$$\mathcal{R}_2 = \min \sqrt{\begin{aligned}&\left(c_{iso}^{target} - c_{iso}\right)^2 + \left(c_{hex}^{target} - c_{hex}\right)^2 \\ &+ \left(c_{tet}^{target} - c_{tet}\right)^2 + \left(c_{ort}^{target} - c_{ort}\right)^2\end{aligned}}, \qquad (8)$$

where $c_{iso}^{target}, c_{hex}^{target}, c_{tet}^{target}, c_{ort}^{target}$ and $c_{iso}, c_{hex}, c_{tet}, c_{ort}$ are the normalized symmetry classes of the target and the parameterized microstructures, respectively; $c_{iso}, c_{hex}, c_{tet}$ and $c_{ort}$ are functions of $t_c, t_h, t_v$ and $t_e$.

For the two approaches, geometric parameters are subjected to the inequality restrictions in equation (6) and at the same time, they must comply with the restriction imposed by the bone volume-to-total volume ratio of the target microstructure,

$$BV/TV(t_c, t_h, t_v, t_e) = BV/TV^{target}, \qquad (9)$$

Alternatively, the restriction of the bone volume-to-total volume ratio might be relaxed, such as

$$\begin{aligned}BV/TV^{target}\left(1 - BV/TV_{tol}\right) &\leq BV/TV(t_c, t_h, t_v, t_e) \\ &\leq BV/TV^{target}\left(1 + BV/TV_{tol}\right),\end{aligned} \qquad (10)$$

where $BV/TV_{tol}$ is a prescribed tolerance that could take any value in the interval [0,1].

## 4.2 Algorithms

Based on the continuity of the interpolations of $\mathbb{C}$ and its symmetry classes (see Section 3.3), different algorithms are used for the optimization problems posed in equations (7) and (8).

Being the interpolation of the $\mathbb{C}$ coefficient continuous, the optimization problem (7) is solved using and active-set Sequential Quadratic Programming (SQP) algorithm. SQP methods solve a sequence of optimization subproblems, each of which optimizes a quadratic model of the objective function subjected to a linearization of the constraints. We use in this work the SQP algorithm of the Matlab Optimization Toolbox (`fmincon`), which is a customized implementation of the algorithm by Gill et al. (1984, 1991). `fmincon` is gradient-based, and it works on problems where the objective and constraint functions are both continuous and have continuous first derivatives. `fmincon` finds the minimum of a problem specified by

$$\min_x f(x) \text{ such that } \begin{cases} c_{eq}(x) = 0 \\ c_{ineq}(x) \leq 0 \\ A_{ineq}(x) \cdot x \leq b_{ineq} \\ b_{low} \leq x \leq b_{up} \end{cases} \quad (11)$$

where $x = [t_c, t_h, t_v, t_e]^T$ and the objective function is $f(x) = \mathcal{R}_1(t_c, t_h, t_v, t_e)$, see equation (7).

The constrain functions are used for the restriction on $BV/TV$. In the case of a prescribed target $BV/TV$, the equality function constrain, $c_{eq}(x) = 0$, is

$$BV/TV(t_c, t_h, t_v, t_e) - BV/TV^{target} = 0, \quad (12)$$

while if $BV/TV$ is defined as in equation (10), two $c_{ineq}(x) \leq 0$ are specified:

$$BV/TV(t_c, t_h, t_v, t_e) - BV/TV^{target}(1 + BV/TV_{tol}) \leq 0 \quad (13)$$

and

$$BV/TV^{target}(1 - BV/TV_{tol}) - BV/TV(t_c, t_h, t_v, t_e) \leq 0. \quad (14)$$

Inequality constrains are the restrictions on the geometric parameters in equation (6). Thus, $A_{ineq}(x) \cdot x \leq b_{ineq}$ is

$$\begin{bmatrix} 1 & -1 & 0 & 0 \\ 1 & 0 & -1 & 0 \end{bmatrix} \begin{bmatrix} t_c \\ t_h \\ t_v \\ t_e \end{bmatrix} \leq \begin{bmatrix} 0 \\ 0 \end{bmatrix}. \quad (15)$$

Finally, inequality constrains $b_{low} \leq x \leq b_{up}$ are used to set the validity intervals for the polynomial interpolations in Section 3.2. For the polynomials of order 10, these are

$$b_{low} = \begin{bmatrix} 0.06 \\ 0.06 \\ 0.06 \\ 0.6 \end{bmatrix} \text{ and } b_{up} = \begin{bmatrix} 0.95 \\ 0.95 \\ 0.95 \\ 1.4 \end{bmatrix}. \quad (16)$$

Stopping criteria consist of two tolerances: `FunctionTolerance`, a lower bound on the change in the value of the objective function during a step, and `StepTolerance`, the termination tolerance on the step size. Iterations ends when either of the conditions is achieved. Both tolerances were set with the default values of $10^{-6}$.

The optimization posed in term of the symmetry classes is solved using the derivative-free constrained direct search solver `patternsearch` (PS) of the Matlab Global Optimization Toolbox. `patternsearch` computes a sequence of points that approach an optimal. At each step, the algorithm searches a set of points, called a mesh, around the current point. The mesh is formed by adding the current point to a scalar multiple of a set of vectors called a pattern. If the pattern search algorithm finds a point in the mesh that improves the objective function at the current point, the new point becomes the current point at the next step of the algorithm.

The problem for the PS algorithm is specified using the same outline for the SQP in equation (11), but with the objective function $f(x) = \mathcal{R}_2(t_c, t_h, t_v, t_e)$ of equation (8). Restrictions to the geometric parameters due to the feasibility of the microstructures, $BV/TV$ and confidence intervals for the polynomial interpolations are the same as for the SQP algorithm. Stopping criteria involves tolerances for `FunctionTolerance`, the difference between the function value at the previous best point and function value at the current best point; `MeshTolerance`, the minimum size for the search mesh; and `StepTolerance`, the minimum distance from the previous best point to the current best point. The three tolerances were set with the default values of $10^{-6}$.

### 4.3 Verification and tuning

The optimization procedures were verified, tested and tuned by assessing its effectiveness to identify microstructures among those of the database used for the elasticity-matrix polynomial fitting (see Section 3.3). To this end, 100 parameterized microstructures were randomly selected from the database to serve as target microstructures. The optimization problems were solved with the SQP and PS methods for $BV/TV_{tol} = 1\%$ and 5%. Since the thermodynamic requirement was checked only for discrete combinations $(t_c, t_h, t_v, t_e)$ in Section 3.3, there is no guarantee that all possible combinations will satisfy it. Therefore, the thermodynamic requirement for $\mathbb{C}(t_c, t_h, t_v, t_e)$ was checked at the end of each optimization procedure.

Preliminary tests had shown that the performances of both algorithms were sensitive to the initial values (seeds) of the geometric parameters. Thus, SQP optimizations were attempted with different seeds as many times as necessary until the objective functions attained the condition $\mathcal{R}_1 < 0.001$; the PS optimizations were run four times for different sets of random seeds and the best result reported.

Table 6 reports the mean values of the residuals and the mean values and standard deviations of the relative errors of the geometric parameters and the symmetry classes. Errors for the geometric parameters are

$$e_{t_c} = \frac{t_c - t_c^{target}}{t_c^{target}}, \qquad e_{t_h} = \frac{t_h - t_h^{target}}{t_h^{target}},$$
$$e_{t_v} = \frac{t_v - t_v^{otarget}}{t_v^{target}} \quad \text{and} \quad e_{t_e} = \frac{t_e - t_e^{target}}{t_e^{target}}. \tag{17}$$

Besides, errors for the symmetry classes are relative to the corresponding average value of the 100 target microstructures. This approach avoids the occurrence of boundless and misleading large errors for the target microstructures with zero or nearly zero symmetry classes. Thus, the errors for the symmetry classes are defined as follows

$$e_{c_{iso}} = \frac{c_{iso} - c_{iso}^{target}}{\bar{c}_{iso}^{target}}, \qquad e_{c_{hex}} = \frac{c_{hex} - c_{hex}^{target}}{\bar{c}_{hex}^{target}}, \tag{18}$$

$$e_{c_{tet}} = \frac{c_{tet} - c_{tet}^{target}}{\overline{c_{tet}^{target}}} \quad \text{and} \quad e_{c_{ort}} = \frac{c_{ort} - c_{ort}^{target}}{\overline{c_{ort}^{target}}},$$

where $\overline{c_{iso}^{target}}$, $\overline{c_{hex}^{target}}$, $\overline{c_{tet}^{target}}$ and $\overline{c_{ort}^{target}}$ are the mean values of the symmetry class fractions for the target microstructures.

| Method | $BV/TV_{tol}$ [%] | Residual | Error geom. parameters [%] | | | | Symmetry-class errors x 10⁻³ | | | |
|---|---|---|---|---|---|---|---|---|---|---|
| | | | $t_c$ | $t_h$ | $t_v$ | $t_e$ | $c_{iso}$ | $c_{hex}$ | $c_{tet}$ | $c_{ort}$ |
| SQP | 0 | 3.1×10⁻⁴ | -1.0 ± 3.9 | 1.1 ± 4.9 | 0.3 ± 1.4 | 0.02 ± 0.09 | 0.08 ± 0.89 | 500 ± 1034 | -916 ± 1036 | -250 ± 761 |
| | 1 | 1.7×10⁻⁴ | -0.4 ± 2.9 | 0.3 ± 2.6 | 0.1 ± 1.1 | 0.00 ± 0.07 | 0.06 ± 0.53 | 499 ± 1035 | -916 ± 1037 | -250 ± 761 |
| | 5 | 2.3×10⁻⁴ | 0.1 ± 7.7 | 0.1 ± 4.2 | -0.1 ± 2.4 | 0.00 ± 0.07 | 0.03 ± 0.71 | 500 ± 1035 | -916 ± 1037 | -250 ± 761 |
| PS | 0 | 1.4×10⁻² | 17 ± 41 | -0 ± 38 | 0 ± 33 | 2 ± 31 | -4.8 ± 18.3 | 7 ± 237 | 80 ± 502 | 14 ± 137 |
| | 1 | 5.6×10⁻³ | 12 ± 33 | 1 ± 34 | -5 ± 18 | 4 ± 25 | 0.7 ± 11.7 | 18 ± 168 | -33 ± 262 | -0.4 ± 47.9 |
| | 5 | 3.1×10⁻³ | 5 ± 37 | 4 ± 36 | 6 ± 36 | 5 ± 26 | -0.8 ± 4.2 | -1.4 ± 88.9 | 27 ± 175 | 10 ± 66 |

*Table 6: Mean values of the residuals and errors for the geometric parameters and the symmetry classes.*

Table 6 shows that the SQP optimization produces the best results for the geometric parameters. SQP errors for the geometric parameters diminish with the relaxation of the $BV/TV_{tol}$, maximum errors are around 1% for $BV/TV_{tol} = 0$ and they reduce to less than 0.1% for $BV/TV_{tol} = 5\%$. Maximum errors are for $t_c$ and $t_h$. Standard deviations are, in general, within a few percent. Symmetry class errors behave almost independently of $BV/TV_{tol}$. The result for $c_{iso}$ is very accurate, but in contrast, errors and standard deviations for $c_{hex}$, $c_{tet}$ and $c_{ort}$ are very large. Mean values of $BV/TV$ for the optimal microstructures are almost coincident to the target values, with almost no effect of $BV/TV_{tol}$; mean values for the relative error

$$e_{BV/TV} = \frac{BV/TV - BV/TV^{target}}{BV/TV^{target}}, \tag{19}$$

are $\overline{e_{BV/TV}} = 6 \cdot 10^{-5}$ and $-5 \cdot 10^{-4}$ for $BV/TV_{tol} = 1\%$ and 5%, respectively.

The PS optimization produces accurate results for the symmetry classes, with maximum errors of a few percent for $e_{tet}$. Error $e_{tet}$ diminishes from 8% to 3% with the increment of $BV/TV_{tol}$. However, this is not the case for all the symmetry classes; note that $e_{hex}$ and $e_{ort}$ present their minima for $BV/TV_{tol} = 1\%$. Maximum standard deviations are also for $e_{tet}$, and they reduce from 50% to 17% with the increment of $BV/TV_{tol}$. The largest error in the geometric parameters is for $t_c$, which diminishes from 17% to 5% as the tolerance for the bone volume fraction is relaxed from $BV/TV_{tol} = 0$ to $BV/TV_{tol} = 5\%$. In contrast, minimum errors for $t_h$, $t_v$ and $t_e$ are for $BV/TV_{tol} = 0$, and they deteriorate with $BV/TV_{tol}$; in any case, maximum errors are of a few percent. The standard deviations for the geometric parameters are not sensitive to $BV/TV_{tol}$ and they are from 25% to 40%. Like for the SQP procedure, mean $BV/TV$ of the optimal microstructures are very close the target values, in these cases $\overline{e_{BV/TV}} = 7 \cdot 10^{-4}$ and $-5 \cdot 10^{-3}$ for $BV/TV_{tol} = 1\%$ and 5%, respectively.

In what respects to the computational performance, SQP algorithm had to be run, in average, 1.4 times for different seeds to attain the goal $\mathcal{R}_1 < 0.001$. Each run comprised an average of 32 iterations and 180 function evaluations. The performance of the SQP algorithm was independent of $BV/TV_{tol}$. The PS algorithm employed on average 3.7 iterations with 1720 function evaluations for each of the seed value sets. It is worth noting that, although the larger number of function evaluations, the PS method was always faster than the SQP. This is because, besides the greater complexity of the algorithm, the SPQ method needs of the cost function gradient, which is around five times more expensive to evaluate than the evaluation of the cost function itself.

The above results show that SQP and PS methods are effective to retrieve randomly selected specimens from the parameterized microstructure database. The SQP method produces accurate results in terms of the geometric parameters, but they are poor in what respects to the symmetry classes. The PS method produces accurate results in terms of the symmetry classes, which, at the same time, show good results for the geometric parameters. The two methods are further assessed next, by finding parameterized microstructures that better mimic the elastic response of natural microstructures.

## 5 APPLICATION TO NATURAL TRABECULAR SAMPLES

### 5.1 Human specimens

Elasticity matrices of the 141 human bone samples were filtered to retrieve their orthotropic parts. To do so, components $C_{14}, C_{15}, C_{16}, C_{24}, C_{25}, C_{26}, C_{34}, C_{35}, C_{36}, C_{45}, C_{46}$ and $C_{56}$ were set equal to zero for the elastic tensors oriented in the symmetry Cartesian coordinate system and their symmetry class decompositions were computed such that $c_{ort}^{obj} + c_{tet}^{obj} + c_{hex}^{obj} + c_{iso}^{obj} = 1$. The resultant data was used as target values for the SQP and PS optimization procedures. Optimizations were performed for $BV/TV_{tol} = 0\%, 1\%, 5\%$ and $10\%$.

Symmetry classes for the obtained effective elastic properties are assessed in relation to the mean values of their extreme fractions for the human microstructures in Table 3, this is:

$$e'_{c_{iso}} = \frac{c_{iso} - c_{iso}^{target}}{\frac{1}{2}\left(c_{iso}^{max} + c_{iso}^{min}\right)}, \qquad e'_{c_{hex}} = \frac{c_{hex} - c_{hex}^{target}}{\frac{1}{2}\left(c_{hex}^{max} + c_{hex}^{min}\right)},$$

$$e'_{c_{tet}} = \frac{c_{tet} - c_{tet}^{target}}{\frac{1}{2}\left(c_{tet}^{max} + c_{tet}^{min}\right)} \text{ and } e'_{c_{ort}} = \frac{c_{ort} - c_{ort}^{target}}{\frac{1}{2}\left(c_{ort}^{max} + c_{ort}^{min}\right)}. \tag{20}$$

Neither the SQP nor the PS showed significant improvements with the relaxation of the $BV/TV_{tol}$ constrain, so only the results for $BV/TV_{tol} = 1\%$ are presented next. Mean errors for 141 samples when using the SQP method are $\overline{e'_{c_{iso}}} = 0.23$, $\overline{e'_{c_{hex}}} = -0.50$, $\overline{e'_{c_{tet}}} = 0.54$ and $\overline{e'_{c_{ort}}} = 0.17$, while for the PS method they are $\overline{e'_{c_{iso}}} = 0.07$, $\overline{e'_{c_{hex}}} = -0.09$, $\overline{e'_{c_{tet}}} = -0.03$ and $\overline{e'_{c_{ort}}} = -0.02$. It can be observed that, like in Section 4.3, the PS achieved better results than the SQP.

Figure 9 presents the results for the PS optimizations in terms of the relative error for the elasticity matrices,

$$e_{\|\mathbb{C}\|} = \frac{\|\mathbb{C}\| - \|\mathbb{C}^{target}\|}{\|\mathbb{C}^{target}\|}. \tag{21}$$

It can be observed that $\overline{e_{\|\mathbb{C}\|}}$ and its dispersion diminish with $BV/TV$, from $0.3 \lesssim e_{\|\mathbb{C}\|} \lesssim 4$ for $BV/TV < 10\%$ to $0.2 \lesssim e_{\|\mathbb{C}\|} \lesssim 0.8$ for $BV/TV > 32\%$. At the same time, it is interesting to note that, with only a few exceptions, $e_{\|\mathbb{C}\|} > 0$, what implies that the optimized parameterized microstructures are, in general, stiffer than the target natural ones. It might be argued that the parameterized microstructures make, in terms of stiffness, a more efficient use of the material than natural microstructures, being this greater efficiency more noticeable for low $BV/TV$. This behavior allows to explain the poor performance of the SQP optimization: since parameterized microstructures are in general stiffer than the natural microstructures for a given $BV/TV$, the objective function posed in terms of the elastic matrices (see (7)) is hard to minimize. On the other hand, the PS optimization, which is posed in terms of the symmetry classes, is not affected by the overall stiffness, and thus it conducts to better results. The $BV/TV$ results confirm this observation: mean values of the error in (19) are close to zero for the PS method irrespectively of $BV/TV_{tol}$ ($\overline{e_{BV/TV}} = -1 \cdot 10^{-3}$, $5 \cdot 10^{-3}$ and $2 \cdot 10^{-2}$ for $BV/TV_{tol} = 1\%, 5\%$ and $10\%$, respectively), whereas for the SQP they are always almost coincident with the lowest value of the tolerance range.

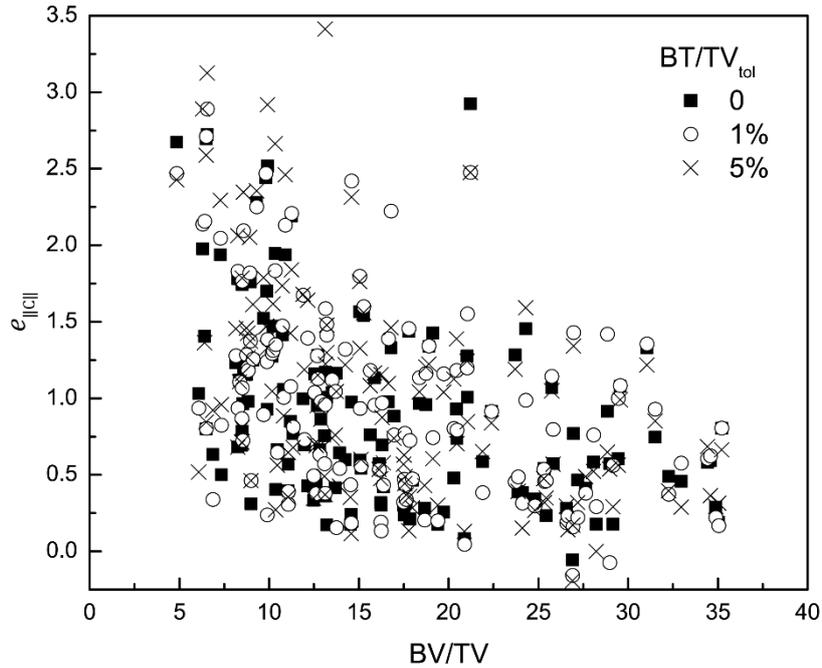

*Figure 9: PS optimization of the human samples: stiffness-matrix error as function of BV / TV.*

The solution of the PS optimization can be improved for $e_{\|\mathbb{C}\|}$ by doing a convenient selection of the Young's modulus of the artificial microstructure material. If the artificial microstructure is constructed using a material with Young´s modulus $E' = \frac{\|\mathbb{C}^{target}\|}{\|\mathbb{C}\|} E$, the error for the elastic matrix norm $e_{\|\mathbb{C}\|} = 0$. Clearly, this scaling of the Young´s modulus does not affect the elastic symmetries. The above analysis was performed for the 141 samples. The average scaling factor for the Young´s modulus was found $\overline{E'/E} = 0.55$ with and standard deviation $SD_{E'/E} = 0.20$. Resultant mean errors and standard deviations for the elasticity coefficients are reported in Table 7. Errors are reported relative to the individual elasticity coefficients,

$$e_{ij}^{C} = \frac{C_{ij} - C_{ij}^{target}}{C_{ij}^{target}}, \tag{22}$$

and relative to the elasticity matrix norm

$$e_{ij}^{\|\mathbb{C}\|} = \frac{C_{ij} - C_{ij}^{target}}{\|\mathbb{C}^{target}\|}. \tag{23}$$

It is observed that coefficients $C_{12}$, $C_{13}$ and $C_{11}$ have the maximum individual relative errors, which range from 80% to 210%; however, when evaluated relative to $\|\mathbb{C}^{target}\|$, maximum errors do not exceed 10%.

| Error | | Relative errors for elasticity matrix coefficients | | | | | | | | |
|---|---|---|---|---|---|---|---|---|---|---|
| | | $C_{11}$ | $C_{22}$ | $C_{33}$ | $C_{23}$ | $C_{13}$ | $C_{12}$ | $C_{44}$ | $C_{55}$ | $C_{66}$ |
| $e_{ij}^{C}$ | mean | 0.80 | -0.04 | -0.24 | 0.34 | 0.86 | 2.15 | -0.28 | 0.05 | -0.21 |
| | SD | 1.06 | 0.34 | 0.26 | 0.66 | 0.83 | 1.49 | 0.38 | 0.54 | 0.81 |
| $e_{ij}^{\|\mathbb{C}\|}$ | mean | 0.10 | -0.02 | -0.11 | 0.01 | 0.04 | 0.08 | -0.04 | 0.00 | -0.01 |
| | SD | 0.13 | 0.08 | 0.14 | 0.03 | 0.03 | 0.03 | 0.04 | 0.03 | 0.04 |

*Table 7: Relative errors for the elasticity matrix after the PS optimization.*

Further results concerning the performance of the SQP and PS algorithms are given in the Appendix.

## 5.2 Bovine specimens

The PS optimization was performed for the orthotropic part of the stiffness matrices of the 5 bovine specimens described in Section 2.1. Tolerance for the volume fraction was set $BV/TV_{tol} = 1\%$.

The optimized microstructures are shown in Figure 10 together with their corresponding target natural samples. The stiffness matrices of the parameterized microstructures are given in the Appendix; their geometric parameters and $BV/TV$ values are reported in Table 8. Table 9 and Table 10 report the errors for the symmetry classes and the elasticity coefficients, respectively. Error for the symmetry classes are in relation to the ranges of the extreme fractions for the bovine microstructures in Table 3 as in equation (20).

Table 9 shows that, with the only exception of the tetragonal symmetry, errors for the symmetry classes are very low. The large relative errors for the tetragonal symmetry are explained due to its small relative contribution to the elasticity matrix (see Figure 2). The worst performance is for Sample #2, which has the particularity of having the lowest isotropic class and the highest orthorhombic class fractions, $c_{iso} = 0.54$ and $c_{ort} = 0.27$, respectively. Bone volume fractions of the mimetic microstructures are coincident to those of the natural specimens (see Table 1) up to the second significant figure, which is coherent to the $BV/TV_{tol} = 1\%$ used for the analyses.

Error for the elasticity coefficients in Table 10 were computed after the scaling of the Young´s modulus to make $e_{\|\mathbb{C}\|} = 0$. The resultant scaling factors range $0.41 < E'/E < 0.72$ with a mean value of $\overline{E'/E} = 0.55$, which coincides with that of the human sample analysis in the previous section. It is observed that as for the human samples, $C_{12}$ presents the highest mean error level (around 9%); maximum errors are of around 17% for $C_{11}$ and $C_{22}$ of Sample #5.

*Figure 10: Natural bovine specimens and their mimetic parameterized microstructures.*

| Sample | Geometric parameters | | | | $BV/TV$ [%] |
|---|---|---|---|---|---|
| | $t_c$ | $t_h$ | $t_v$ | $t_e$ | |
| 1 | 0.080 | 0.674 | 0.949 | 0.763 | 25 |
| 2 | 0.187 | 0.229 | 0.931 | 0.600 | 38 |
| 3 | 0.217 | 0.217 | 0.521 | 0.890 | 30 |
| 4 | 0.080 | 0.391 | 0.718 | 0.727 | 20 |
| 5 | 0.100 | 0.476 | 0.497 | 1.256 | 21 |

*Table 8: Geometric parameters and $BV/TV$ of the mimetic bovine samples.*

| Sample | Errors for symmetry classes | | | |
|---|---|---|---|---|
| | $e'_{c_{iso}}$ | $e'_{c_{hex}}$ | $e'_{c_{tet}}$ | $e'_{c_{ort}}$ |
| 1 | 0.002 | 0.011 | -0.076 | 0.018 |
| 2 | 0.024 | 0.166 | 0.332 | -0.696 |
| 3 | 0.001 | 0.006 | -0.049 | 0.013 |
| 4 | -0.002 | -0.010 | 0.074 | -0.019 |
| 5 | -0.001 | -0.006 | 0.039 | -0.004 |

*Table 9: Errors for the symmetry classes of the mimetic bovine samples.*

| Sample | Relative errors for elasticity matrix coefficients | | | | | | | | |
|---|---|---|---|---|---|---|---|---|---|
| | $C_{11}$ | $C_{22}$ | $C_{33}$ | $C_{23}$ | $C_{13}$ | $C_{12}$ | $C_{44}$ | $C_{55}$ | $C_{66}$ |
| 1 | 0.034 | 0.016 | -0.008 | -0.039 | -0.047 | 0.118 | -0.028 | -0.082 | -0.032 |
| 2 | 0.060 | -0.016 | -0.018 | 0.007 | 0.012 | 0.065 | -0.007 | 0.015 | -0.072 |
| 3 | -0.004 | 0.041 | -0.035 | 0.005 | 0.003 | 0.048 | -0.016 | 0.009 | -0.054 |
| 4 | 0.020 | 0.034 | -0.048 | 0.029 | 0.014 | 0.087 | -0.023 | -0.001 | -0.092 |
| 5 | 0.167 | -0.168 | -0.039 | -0.101 | 0.047 | 0.143 | -0.062 | 0.003 | -0.009 |

*Table 10: Relative errors for the elasticity matrix of the mimetic bovine samples.*

# 6   CONCLUSIONS

This work introduces a procedure for the design of artificial parameterized microstructures that mimic the elastic response of cancellous bone. The procedure is based on the parameterized microstructure by Kowalczyk (2006), the geometric parameters of which are optimized to minimize the differences between the symmetry classes of the target and the artificial microstructure elastic tensor.

Symmetry class analyses of experimental data from Kabel et al. (1999a, 1999b) and of specimens processed as part of this work show that elastic symmetries can be related to the specimen $BV/TV$. The isotropic symmetry class constitutes the main fraction of the elastic tensor; it increases linearly with the specimen bone volume fraction, from around 50% for $BV/TV = 5\%$ to 70% for $BV/TV = 35\%$. The isotropic and hexagonal classes add to a constant, such that they account for around 82% of the elastic tensor over the complete $BV/TV$ range. The orthotropic symmetry, given by the addition of the isotropic, hexagonal, tetragonal and orthorhombic classes, constitute around 93% of the elastic tensor, independently of the $BV/TV$.

The parameterized artificial microstructure is orthotropic by construction. It is shown in this work that it has the capability to combine the isotropic, hexagonal, tetragonal and orthorhombic symmetry classes in the proportions present in the cancellous bone. Analytical expressions for the elastic matrix in terms of the microstructure geometrical parameters are provided. These expressions could be integrated into multiscale design methodologies and used to explore the design of bone substitutes and natural micro-scaffolds. Free Material Optimization Methods (FMO) are seen as a promising approach in this sense.

Two optimization methods are proposed to find the parameterized microstructure that better mimics the elastic response of a target natural bone specimen: a Sequential Quadratic Programming algorithm to minimize the difference between the elasticity matrices, and a Pattern Search algorithm to minimize the difference between the symmetry class decompositions. Both approaches use the geometry parameters as design variables, polynomial interpolations to evaluate the parameterized microstructure elasticity matrix, and $BV/TV$ as a restriction. The Pattern Search approach is found to produce the best results. The analyses of 146 natural cancellous bone specimens resulted in mimetic microstructures whose symmetry class decompositions differ on average 6% with respect to the target values.

The results for the elasticity matrix error allows to observe that the optimized microstructures are in general stiffer than their natural counterparts; this behavior is more noticeable for low $BV/TV$. This deviation can be compensated by selecting the Young's modulus for the optimized microstructure material such that norm of the difference between elasticity matrices of the target and optimized microstructure vanishes. Clearly, the Young´s modulus scaling does not affect the elastic symmetries. The mean value for such scaling factor was found equal to 0.55, i.e., the parameterized microstructure material should have, in average, half the stiffness of the trabecular bone tissue. After scaling, average errors between the optimized and target elasticity matrix coefficients do not exceed 10% relative to the matrix norm.

Additional research is needed to further assess the effectivity of the parameterized microstructures to mimic the behavior of cancellous bone elastic response. A promising approach is to compare the elastodynamic responses of natural and parameterized microstructures by means of ultrasound analyses, and to correlate their behaviors with the elastic symmetry classes and the geometrical parameters. In turn, these results could be used to refine the optimization criteria of the Pattern Search method.

# 7    ACKNOWLEDGEMENTS

This work has been supported by projects PIRSES-GA2009_246977 "Numerical Simulation in Technical Sciences" of the Marie Curie Actions FP7-PEOPLE-2009-IRSES of the European Union and by the PICS project "Modeling and Simulation in Multidisciplinary Engineering" MoSiMe funded in the framework of the CAFCI call by CONICET (Argentina) and CNRS (France). This project has received funding from the European Research Council (ERC) under the European Union's Horizon 2020 research and innovation program (grant agreement No 682001, project ERC Consolidator Grant 2015 BoneImplant).

# 8    CONFLICT OF INTEREST

The authors declare that they have no conflict of interest.

# APPENDIX

**Human Samples**

**Erreur ! Source du renvoi introuvable.** depicts the residual $\mathcal{R}_1$ and the symmetry class errors (18) of the SQP optimizations as functions of $BV/TV$. Figure 11($a$) shows that $\mathcal{R}_1$ and its dispersion diminish sharply with $BV/TV$, from $0.5 \lesssim \mathcal{R}_1 \lesssim 3$ for $BV/TV < 7.5\%$ to $0.1 \lesssim \mathcal{R}_1 \lesssim 0.6$ for $BV/TV > 32\%$. The relaxation of the $BV/TV_{tol}$ constrain does not result in significant improvements in $\mathcal{R}_1$ (results not reported here for analyses performed for $BV/TV_{tol} = 10\%$ exhibit the same behavior). Results in **Erreur ! Source du renvoi introuvable.**($b$) allow to observe that $c_{iso}$ and $c_{hex}$ are systematically over and underestimated, respectively; their mean errors are $\overline{e_{c_{iso}}} = 0.20$ and $\overline{e_{c_{hex}}} = -0.74$. Errors tend to decrease with $BV/TV$, the only exception is that of the tetragonal symmetry, which can attain values over the range $-5 < e_{c_{tet}} < 12$. Mean relative errors for the tetragonal and the orthorhombic symmetry classes are $\overline{e_{c_{tet}}} = 1.74$ and $\overline{e_{c_{ort}}} = 0.18$, respectively.

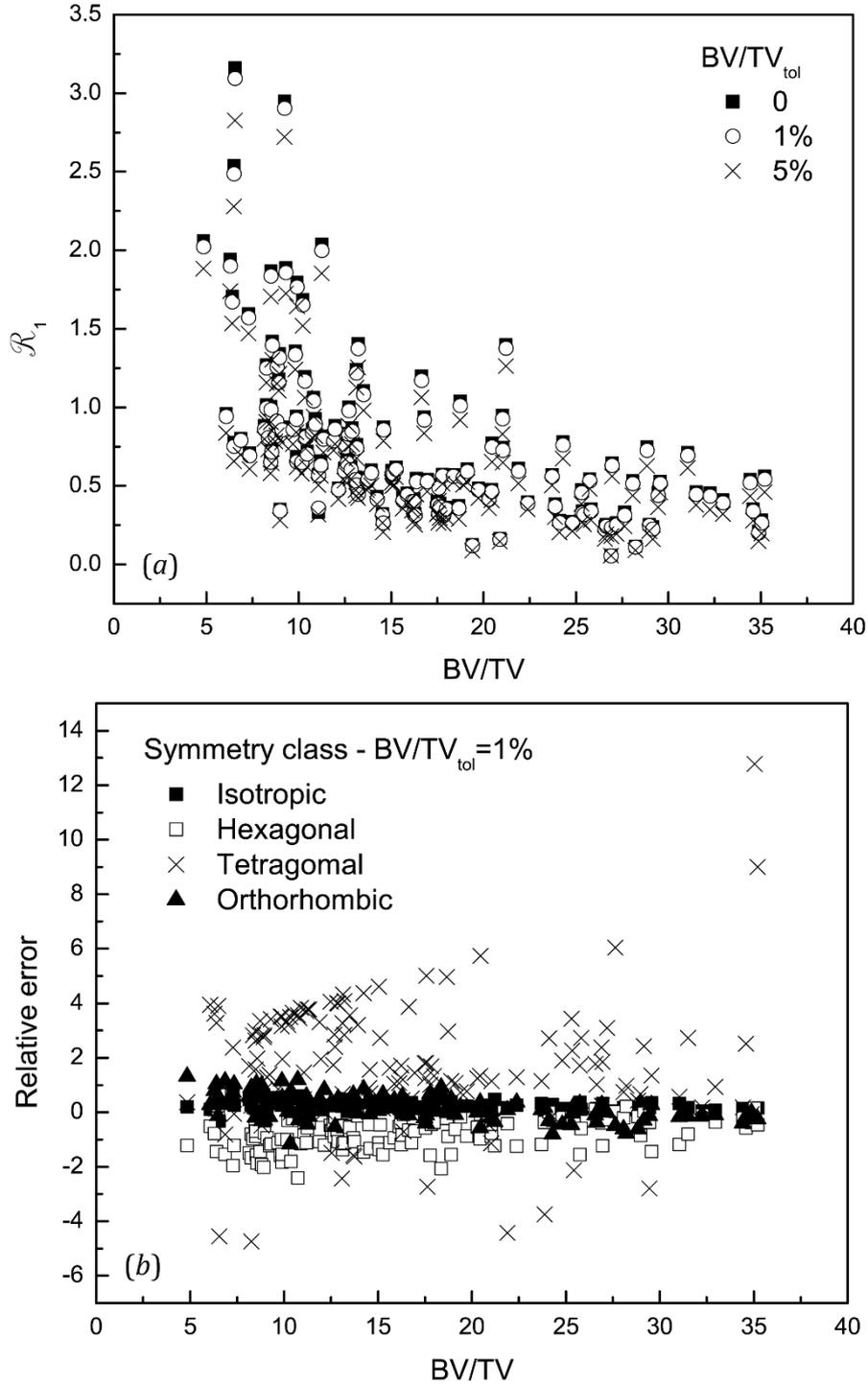

*Figure 11: SQP optimization of human samples: (a) residuals and (b) symmetry-class errors as functions of BV / TV.*

The PS optimization resulted in a mean value for the residual $\overline{\mathcal{R}_2} = 0.06$ with standard deviation $SD_{\mathcal{R}_2} = 0.09$. Like for the SQP approach, the relaxation of the $BV/TV_{tol}$ constrain did not result in significant improvements for $\mathcal{R}_2$. In addition to the error for the elasticity matrices shown in Figure 9, Figure 12 presents the errors (18) for the symmetry classes. Figure 12 allows to observe that mean errors for the symmetry classes are much lower than those of the SQP approach in Figure 11(b) : $\overline{e_{c_{iso}}} = 0.06$, $\overline{e_{c_{hex}}} = -0.13$, $\overline{e_{c_{tet}}} = -0.06$ and $\overline{e_{c_{ort}}} = -0.02$; standard deviations are $SD_{e_{iso}} = 0.11$, $SD_{e_{hex}} = $

0.27, $SD_{e_{hex}} = 0.27$ and $SD_{ort} = 0.18$. Like for the SQP optimization, $e_{c_{tet}}$ presents the largest dispersion and there are tendencies to overestimate $c_{iso}$ and to underestimate $c_{hex}$. The high dispersion of $e_{c_{tet}}$ is consequence of the small relative contribution of the tetragonal symmetry to the elasticity matrix, which is always $c_{tet} < 0.05$, see Table 3 and Figure 1.

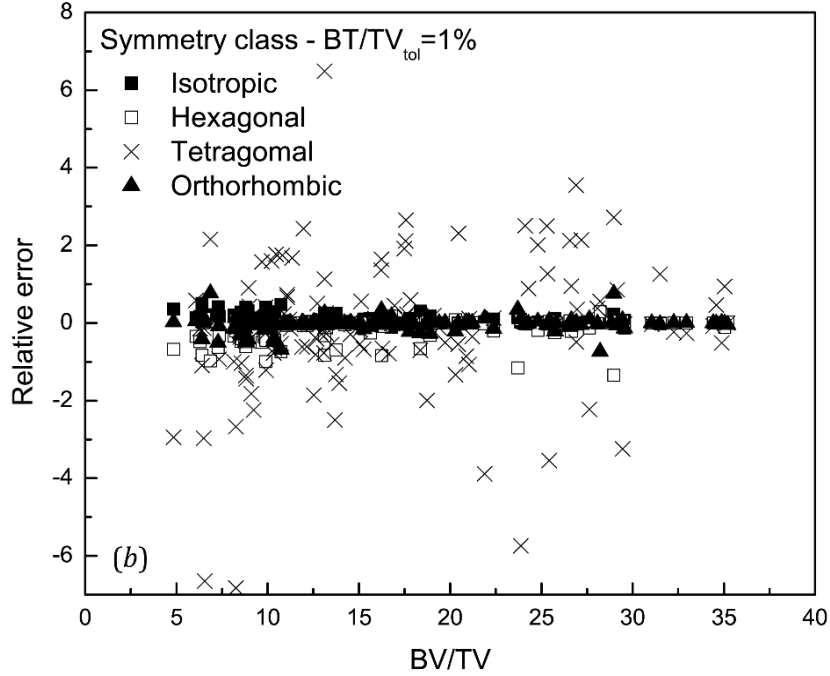

*Figure 12: PS optimization of human samples: symmetry-class error as function of BV / TV.*

**Bovine Samples**

Target elastic matrices of the bovine femoral samples

$$\mathbb{C}_{b_1} = \begin{bmatrix} 0.268 & 0.087 & 0.152 & -0.015 & -0.011 & -0.003 \\ 0.087 & 0.486 & 0.195 & 0.001 & -0.066 & 0.023 \\ 0.152 & 0.195 & 0.677 & 0.010 & 0.044 & 0.011 \\ -0.015 & 0.001 & 0.010 & 0.175 & -0.042 & -0.005 \\ -0.011 & -0.066 & 0.044 & -0.042 & 0.198 & 0.001 \\ -0.003 & 0.023 & 0.011 & -0.005 & 0.001 & 0.112 \end{bmatrix} [\text{GPa}] \quad (\text{A. 1})$$

$$\mathbb{C}_{b_2} = \begin{bmatrix} 0.306 & 0.152 & 0.199 & 0.060 & -0.010 & -0.010 \\ 0.152 & 1.202 & 0.421 & -0.025 & 0.039 & 0.052 \\ 0.199 & 0.421 & 1.885 & -0.040 & 0.067 & 0.007 \\ 0.060 & -0.025 & -0.040 & 0.461 & -0.089 & -0.149 \\ -0.010 & 0.039 & 0.067 & -0.089 & 0.268 & 0.064 \\ -0.010 & 0.052 & 0.007 & -0.149 & 0.064 & 0.256 \end{bmatrix} [\text{GPa}] \quad (\text{A. 2})$$

$$\mathbb{C}_{b_3} = \begin{bmatrix} 0.573 & 0.234 & 0.219 & -0.006 & -0.015 & 0.004 \\ 0.234 & 0.640 & 0.253 & -0.005 & -0.016 & 0.006 \\ 0.219 & 0.253 & 1.260 & -0.017 & 0.023 & -0.003 \\ -0.006 & -0.005 & -0.017 & 0.317 & -0.017 & -0.002 \\ -0.015 & -0.016 & 0.023 & -0.017 & 0.246 & 0.052 \\ 0.004 & 0.006 & -0.003 & -0.002 & 0.052 & 0.224 \end{bmatrix} [\text{GPa}] \quad (\text{A. 3})$$

$$\mathbb{C}_{b_4} = \begin{bmatrix} 0.163 & 0.075 & 0.068 & -0.008 & -0.016 & 0.010 \\ 0.075 & 0.350 & 0.126 & -0.018 & -0.005 & 0.018 \\ 0.068 & 0.126 & 0.606 & 0.029 & -0.004 & -0.032 \\ -0.008 & -0.018 & 0.029 & 0.148 & -0.023 & 0.043 \\ -0.016 & -0.005 & -0.004 & -0.023 & 0.093 & -0.014 \\ 0.010 & 0.018 & -0.032 & 0.043 & -0.014 & 0.110 \end{bmatrix} [\text{GPa}] \quad (\text{A. 4})$$

$$\mathbb{C}_{b_5} = \begin{bmatrix} 0.318 & 0.075 & 0.107 & -0.008 & 0.001 & 0.021 \\ 0.075 & 0.426 & 0.182 & 0.024 & 0.002 & -0.017 \\ 0.107 & 0.182 & 0.464 & -0.027 & -0.003 & -0.001 \\ -0.008 & 0.024 & -0.027 & 0.144 & -0.007 & 0.002 \\ 0.001 & 0.002 & -0.003 & -0.007 & 0.116 & 0.008 \\ 0.021 & -0.017 & -0.001 & 0.002 & 0.008 & 0.114 \end{bmatrix} [\text{GPa}] \quad (\text{A. 5})$$

The PS optimization of each sample was run ten times using different seeds. For each case, the best two outcomes (these are, the solutions that achieve the lowest values of the objective function) were compared to assess the repeatability of the solutions. It was found that for Samples #2 and #4, the residuals of the best two solutions are almost coincident (they differ less than 1%), while the resultant values for the geometrical parameters coincide within 2%. For samples #1 and #3 the lowest two residuals differ in around 12% and the geometrical parameters have discrepancies of up to 80%. It is interesting to note that the parameter $t_e$, the one which governs microstructure orthotropy, showed a remarkable repeatability, it presented discrepancies within 0.2% for the analyses of Samples #1, #2, #3 and #4. In contrast, for Sample #5, residuals of the best two solutions differ in nearly 90% and the geometrical parameters up to 70%; the best performance is for $t_e$, which presents a discrepancy discrepancy of 13%. The behavior for $t_e$ could be explored as a mean for the refinement of the optimization procedure.

Finally, the resultant elastic matrices for the mimetic parameterized microstructures are:

$$\mathbb{C}'_{b_1} = \begin{bmatrix} 0.302 & 0.204 & 0.105 & 0 & 0 & 0 \\ 0.204 & 0.502 & 0.156 & 0 & 0 & 0 \\ 0.105 & 0.156 & 0.669 & 0 & 0 & 0 \\ 0 & 0 & 0 & 0.147 & 0 & 0 \\ 0 & 0 & 0 & 0 & 0.116 & 0 \\ 0 & 0 & 0 & 0 & 0 & 0.081 \end{bmatrix} [\text{GPa}] \quad (\text{B. 1})$$

$$\mathbb{C}'_{b_2} = \begin{bmatrix} 0.268 & 0.087 & 0.152 & 0 & 0 & 0 \\ 0.087 & 0.486 & 0.195 & 0 & 0 & 0 \\ 0.152 & 0.195 & 0.677 & 0 & 0 & 0 \\ 0 & 0 & 0 & 0.175 & 0 & 0 \\ 0 & 0 & 0 & 0 & 0.198 & 0 \\ 0 & 0 & 0 & 0 & 0 & 0.112 \end{bmatrix} [\text{GPa}] \quad (\text{B. 2})$$

$$\mathbb{C}'_{b_3} = \begin{bmatrix} 0.566 & 0.316 & 0.224 & 0 & 0 & 0 \\ 0.316 & 0.709 & 0.261 & 0 & 0 & 0 \\ 0.224 & 0.261 & 1.201 & 0 & 0 & 0 \\ 0 & 0 & 0 & 0.290 & 0 & 0 \\ 0 & 0 & 0 & 0 & 0.261 & 0 \\ 0 & 0 & 0 & 0 & 0 & 0.132 \end{bmatrix} [\text{GPa}] \quad (\text{B. 3})$$

$$\mathbb{C}'_{b_4} = \begin{bmatrix} 0.179 & 0.163 & 0.079 & 0 & 0 & 0 \\ 0.143 & 0.377 & 0.133 & 0 & 0 & 0 \\ 0.079 & 0.133 & 0.569 & 0 & 0 & 0 \\ 0 & 0 & 0 & 0.125 & 0 & 0 \\ 0 & 0 & 0 & 0 & 0.092 & 0 \\ 0 & 0 & 0 & 0 & 0 & 0.038 \end{bmatrix} \text{[GPa]} \quad (B.4)$$

$$\mathbb{C}'_{b_5} = \begin{bmatrix} 0.452 & 0.190 & 0.145 & 0 & 0 & 0 \\ 0.190 & 0.291 & 0.101 & 0 & 0 & 0 \\ 0.145 & 0.101 & 0.433 & 0 & 0 & 0 \\ 0 & 0 & 0 & 0.094 & 0 & 0 \\ 0 & 0 & 0 & 0 & 0.118 & 0 \\ 0 & 0 & 0 & 0 & 0 & 0.106 \end{bmatrix} \text{[GPa]} \quad (B.5)$$